\documentclass[a4paper]{jpconf}

\usepackage{graphicx}
\usepackage{amsmath}
\usepackage{amssymb}
\usepackage{mathrsfs}

\newcommand{\bra}[1]{{\left\langle #1 \right|}} % definition for bra
\newcommand{\ket}[1]{{\left| #1 \right\rangle}} % definition for ket

\begin{document}

\title{Lattice investigation of heavy meson interactions}

\author{Bj\"orn Wagenbach$^1$, Pedro Bicudo$^2$, Marc Wagner$^{1,3}$}

\address{$^1$ Goethe-Universit\"at Frankfurt am Main, Institut f\"ur Theoretische Physik, \\ $\phantom{xxx}$ Max-von-Laue-Stra\ss{}e 1, D-60438 Frankfurt am Main, Germany}
\address{$^2$ Dep.\ F\'{\i}sica and CFTP, Instituto Superior T\'ecnico, Av.\ Rovisco Pais, 1049-001 Lisboa, \\ $\phantom{xxx}$ Portugal}
\address{$^3$ European Twisted Mass Collaboration (ETMC)}

\ead{wagenbach@th.physik.uni-frankfurt.de}

\begin{abstract}
We report on a lattice investigation of heavy meson interactions and of tetraquark candidates with two very heavy quarks. These two quarks are treated in the static limit, while the other two are up, down, strange or charm quarks of finite mass. Various isospin, spin and parity quantum numbers are considered.
\end{abstract}

% ********************
% ********************
% ********************

\section{Introduction}

We study the potential of two static quarks in the presence of two quarks of finite mass. While in \cite{Wagner:2010ad,Wagner:2011ev,Bicudo:2012qt} we have exclusively considered two static antiquarks and two light quarks ($\bar{Q}\bar{Q}ll$), where $l \in \{ u,d \}$, here we also use $s$ and $c$ quarks, i.e.\ investigate $\bar{Q}\bar{Q}ss$ and $\bar{Q}\bar{Q}cc$, to obtain certain insights regarding the quark mass dependence of the static antiquark-antiquark interaction. We also discuss first steps regarding the static quark-antiquark case, i.e.\ $\bar{Q}Q\bar{l}l$, $\bar{Q}Q\bar{s}s$ and $\bar{Q}Q\bar{c}c$.

$\bar{Q}\bar{Q}qq$ systems as well as $\bar{Q}Q\bar{q}q$ systems have been studied also by other groups (cf.\ e.g.\ \cite{Stewart:1998hk,Michael:1999nq,Cook:2002am,Bali:2005fu,Doi:2006kx,Detmold:2007wk,Bali:2010xa,Bali:2011gq,Brown:2012tm}).

% ********************
% ********************
% ********************

\section{Creation operators and trial states}

The $\bar{Q}\bar{Q}qq$ and $\bar{Q}Q\bar{q}q$ potentials $V(r)$ are extracted from correlation functions
\begin{equation}
C(t) \ \ \equiv \ \ \bra{\Omega} \mathcal{O}^\dagger(t) \mathcal{O}(0) \ket{\Omega}
\end{equation} 
according to
\begin{equation}
V(r) \ \ =_{\textrm{large }t} \ \ V_{\textrm{eff}}(r,t) \quad , \quad V_{\textrm{eff}}(r,t) \ \ \equiv \ \ \frac{1}{a} \ln\bigg(\frac{C(t)}{C(t+a)}\bigg) ,
\end{equation} 
where $a$ is the lattice spacing and $\mathcal{O}$ denote suitable creation operators, which are discussed in detail below. For an introduction to lattice hadron spectroscopy cf.\ e.g.\ \cite{Weber:2013eba}.

% ********************

\subsection{Static-light mesons (``$B$ and $\bar{B}$ mesons'')}

The starting point are static-light mesons, which either consist of a static quark $Q$ and an antiquark $\bar{q}$ or of a static antiquark $\bar{Q}$ and a quark $q$ with $q \in \{u,d,s,c\}$. These mesons can be labeled by parity $\mathscr{P} = \pm$, by the $z$-component of the light quark spin $j_z=\pm 1/2$ ($j=1/2$, because we do not consider gluonic excitations) and in case of $q \in \{u,d\}$ by the $z$-component of isospin $I_z = \pm 1/2$ ($I = 1/2$). The lightest static-light meson has $\mathscr{P} = -$ and is commonly denoted by $S$, its heavier parity partner with $\mathscr{P} = +$ by $P_-$. The static-light meson $S$ is an approximation for $B/B^*$, $B_s/B_s^*$ and $B_c$ listed in \cite{PDG}.

We use static-light meson trial states
\begin{equation}
\mathcal{O} \ket{\Omega} \ \ \equiv \ \ \bar{Q} \Gamma q \ket{\Omega}
\end{equation}
with $\Gamma \in \{\gamma_5,\gamma_0\gamma_5,\gamma_j,\gamma_0\gamma_j\}$ for the $S$ and $\Gamma \in \{1,\gamma_0,\gamma_j\gamma_5,\gamma_0\gamma_j\gamma_5\}$ for the $P_-$ meson. For a more detailed discussion of static-light mesons cf.\ \cite{Jansen:2008si,Michael:2010aa}.

% ********************

\subsection{$B \bar{B}$ systems}

We are interested in the potential of two static-light mesons, i.e.\ their energy as a function of their separation $r$. W.l.o.g.\ we separate the mesons along the $z$-axis, i.e.\ their static antiquark $\bar{Q}$ and quark $Q$ are located at $\vec{r}_1 = (0,0,+r/2)$ and $\vec{r}_2 = (0,0,-r/2)$, respectively. The corresponding $B \bar{B}$ trial states are
\begin{equation}
\label{eq_BBbartrial} \mathcal{O} \ket{\Omega} \ \ \equiv \ \ \Gamma_{AB} \tilde{\Gamma}_{CD} \Big(\bar{Q}_C^a(\vec{r}_1) q_A^{(f_1)a}(\vec{r}_1)\Big) \Big(\bar{q}_B^{(f_2)b}(\vec{r}_2) Q_D^b(\vec{r}_2)\Big) \ket{\Omega}
\end{equation}
($A,B,\ldots$ are spin indices, $a,b$ color indices and $(f_1),(f_2)$ flavor indices). Since there are no interactions involving the static quark spins, one should not couple static spins and spins of finite mass, but contract the static spin indices with $\tilde{\Gamma} \in \{ \gamma_5, \gamma_0\gamma_5, \gamma_3, \gamma_0\gamma_3, \gamma_1, \gamma_2, \gamma_0\gamma_1, \gamma_0\gamma_2 \}$. This results in a non-vanishing correlation function independent of $\tilde{\Gamma}$.

The separation of the static quark and the static antiquark restricts rotational symmetry to rotations around the axis of separation, i.e.\ the $z$-axis. Therefore, and since there are no interactions involving the static quark spins, we can label states by the $z$-component of the light quark spin $j_z = -1,0,+1$. For $j_z=0$, i.e.\ for rotationally invariant states, spatial reflections along an axis perpendicular to the axis of separation are also a symmetry operation (w.l.o.g.\ we choose the $x$-axis). The corresponding quantum number is $\mathscr{P}_x = \pm$. $\mathscr{P}_x$ can be used as a quantum number also for $j_z \neq 0$ states, if we use $|j_z|$ instead of $j_z$. Parity $\mathscr{P}$ is not a symmetry, since it exchanges the positions of the static quark and the static antiquark. However, parity combined with charge conjugation, $\mathscr{P} \circ C$ is a symmetry and, therefore, a quantum number. When $q,\bar{q} \in \{u,d\}$, isospin $I \in \{0,1\}$ and its $z$-component $I_z \in \{-1,0,+1\}$ are also quantum numbers. In summary, there are up to five quantum numbers, which label $B\bar{B}$ states, $(I, I_z, |j_z|, \mathscr{P} \circ C, \mathscr{P}_x)$.

% ********************

\subsection{$B B$ systems (and $\bar{B} \bar{B}$ systems)}

We use $B B$ trial states
\begin{equation}
\label{eq_BBtrial} \mathcal{O} \ket{\Omega} \ \ \equiv \ \ (\mathcal{C}\Gamma)_{AB} \tilde{\Gamma}_{CD} \Big(\bar{Q}_C^a(\vec{r}_1) \psi_A^{(f_1)a}(\vec{r}_1)\Big) \Big(\bar{Q}_D^b(\vec{r}_2) \psi_B^{(f_2)b}(\vec{r}_2)\Big) \ket{\Omega}
\end{equation}
with $\tilde{\Gamma} \in \{1, \gamma_0, \gamma_3\gamma_5, \gamma_1\gamma_2, \gamma_1\gamma_5, \gamma_2\gamma_5, \gamma_2\gamma_3, \gamma_1\gamma_3 \}$ ($\mathcal{C} \equiv \gamma_0 \gamma_2$ denotes the charge conjugation matrix). Arguments similar to those of the previous subsection lead to quantum numbers $(I, I_z, |j_z|, \mathscr{P}, \mathscr{P}_x)$. For a more detailed discussion cf.\ \cite{Wagner:2010ad,Wagner:2011ev}.

% ********************
% ********************
% ********************

\section{\label{sec_setup}Lattice setup}

We use three ensembles of gauge link configurations generated by the European Twisted Mass Collaboration (ETMC) (cf.\ Table~\ref{tab_ensembles}). For the $\bar{Q}\bar{Q}qq$ potentials we use $N_f = 2$ ensembles with lattice spacing $a \approx 0.079 \, \textrm{fm}$ for $q \in \{ u,d \}$ and an even finer lattice spacing $a \approx 0.042 \, \textrm{fm}$ for $q \in \{ s,c \}$, because in the latter case the potentials are quite narrow. Existing $\bar{Q}Q\bar{q}q$ results are rather preliminary and have been obtained exclusively with $q = c$ and the $N_f = 2+1+1$ ensemble with $a \approx 0.086 \, \textrm{fm}$. For details regarding these ETMC gauge link ensembles cf.\ \cite{Boucaud:2008xu,Baron:2009wt,Baron:2010bv,Jansen:2011vv,Cichy:2012is}.

\begin{table}[htb]
\begin{center}
\begin{tabular}{ccccccccc}
\br
Ensemble & $N_f$ & $\beta$ & $(L/a)^3 \times (T/a)$ & $a\mu_l$ & $a\mu_\sigma$ & $a\mu_\delta$ & a & $m_\pi$ \\
\mr
A40.24 & 2 & 3.90 & $24^3 \times 48$ & 0.00400 & - & - & $0.079 \, \textrm{fm}$ & $340 \, \textrm{MeV}$ \\
E17.32 & 2 & 4.35 & $32^3 \times 64$ & 0.00175 & - & - & $0.042 \, \textrm{fm}$ & $352 \, \textrm{MeV}$ \\
\mr
A40.24 & 2+1+1 & 1.90 & $24^3 \times 48$ & 0.00400 & 0.15 & 0.19 & $0.086 \, \textrm{fm}$ & $332 \, \textrm{MeV}$ \\
\br
\end{tabular}
\end{center}
\caption{\label{tab_ensembles}ETMC gauge link ensembles used in this work.}
\end{table} 

Correlation functions have been computed using around 100 gauge link configurations from each of the three ensembles. We have checked that these correlation functions transform appropriately with respect to the symmetry transformations (1) twisted mass time reversal, (2) twisted mass parity, (3) twisted mass $\gamma_5$-hermiticity, (4) charge conjugation and (5) cubic rotations. In a second step we have averaged correlation functions related by those symmetries to reduce statistical errors.

% ********************
% ********************
% ********************

\section{Numerical results}

% ********************

\subsection{$\bar{Q}\bar{Q}qq$ potentials}

In the following we focus on the attractive channels between ground state static-light mesons ($S$ mesons). For $q \in \{ u,d \}$ there is a more attractive scalar isosinglet ($qq = (ud - du)/\sqrt{2}$, $\Gamma = \gamma_5 + \gamma_0 \gamma_5$ corresponding to quantum numbers $(I, |j_z|, \mathscr{P}, \mathscr{P}_x) = (0,0,-,+)$) and a less attractive vector isotriplet ($qq \in \{ uu,(ud + du)/\sqrt{2},dd \}$, $\Gamma = \gamma_j + \gamma_0 \gamma_j$ corresponding to quantum numbers $(I, |j_z|, \mathscr{P}, \mathscr{P}_x) = (1,\{0,1\},-,\pm)$). For $qq = ss$ there is only a single attractive channel, the equivalent of the vector isotriplet. To study also the scalar isosinglet with $s$ quarks, we consider two quark flavors with the mass of the $s$ quark, i.e.\ $qq = (s_1 s_2 - s_2 s_1)/\sqrt{2}$. Similarly we consider $qq = (c_1 c_2 - c_2 c_1)/\sqrt{2}$ to study a charm scalar isosinglet.

\begin{figure}[htb]
\begin{center}
\includegraphics[width=13.0pc,angle=270]{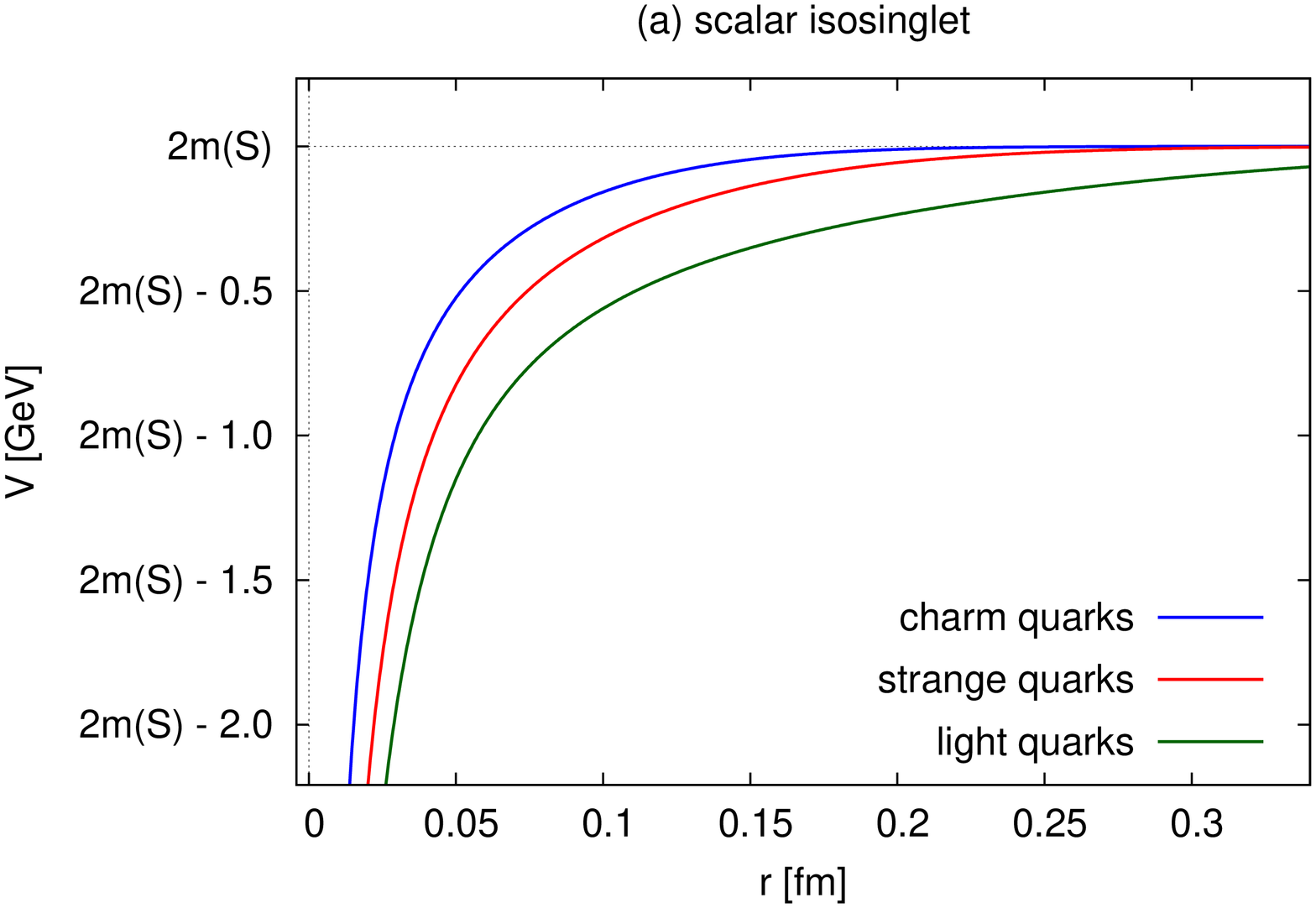}\includegraphics[width=13.0pc,angle=270]{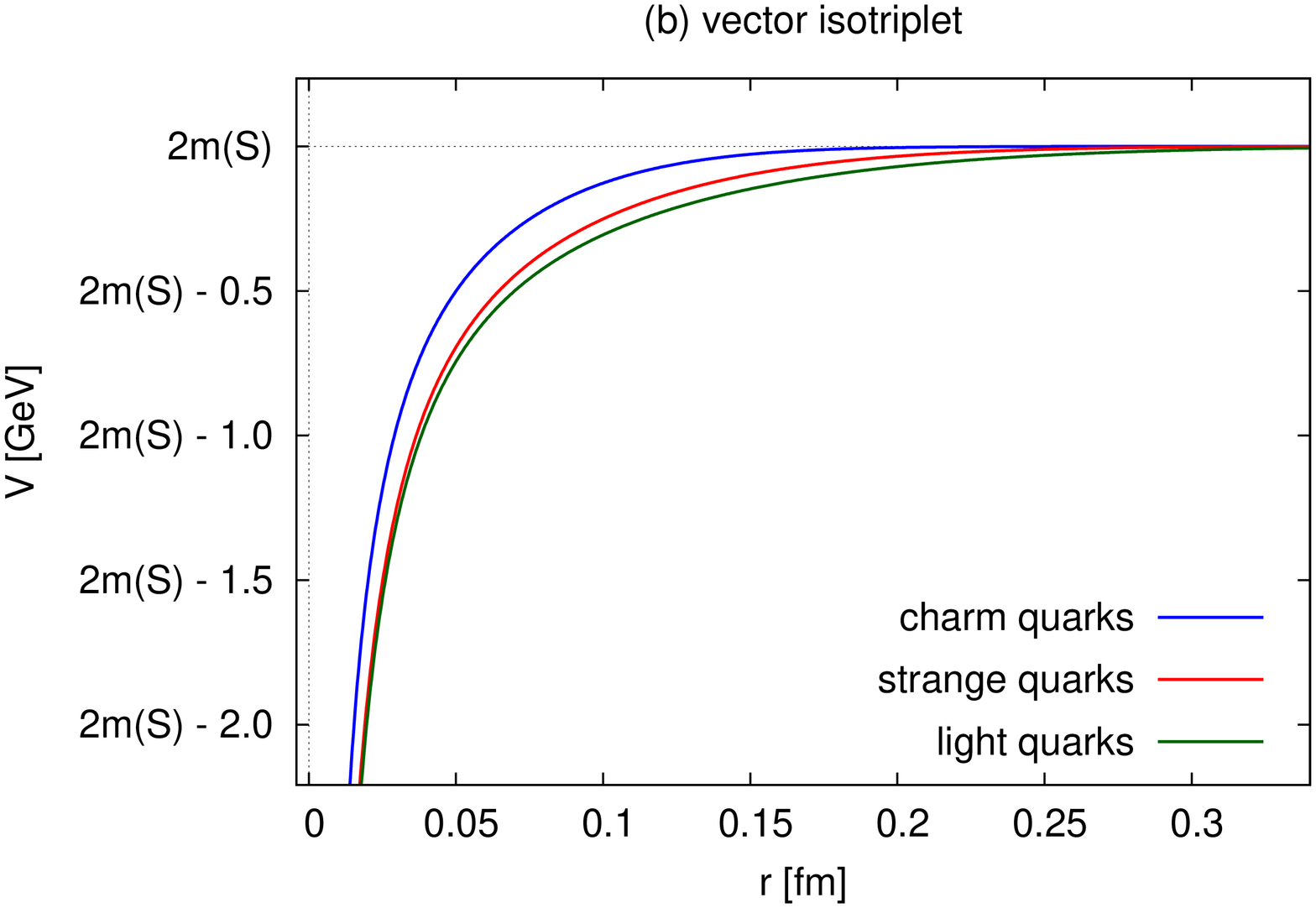}
\end{center}
\caption{\label{FIG001}$\bar{Q}\bar{Q}qq$ potentials (\ref{eq_potfit}) for $q = u/d$, $q = s$ and $q = c$ (error bands are not shown).
\textbf{(a)}~Scalar isosinglet.
\textbf{(b)}~Vector isotriplet.}
\end{figure}

Proceeding as in \cite{Bicudo:2012qt} we perform $\chi^2$ minimizing fits of
\begin{equation}
\label{eq_potfit} V(r) \ \ = \ \ -\frac{\alpha}{r} \exp\bigg(-\bigg(\frac{r}{d}\bigg)^p\bigg)
\end{equation}
with respect to the parameters $d$ (light isotriplet), $(d,\alpha)$ ($q = s$ or $q = c$) or $(d,\alpha,p)$ (light isosinglet) to the lattice results for the $\bar{Q}\bar{Q}qq$ potentials. The resulting functions $V(r)$ are shown in Figure~\ref{FIG001}.

To determine, whether the investigated mesons may form a bound state, i.e.\ a tetraquark, we insert the potentials shown in Figure~\ref{FIG001} into Schr\"odinger's equation with reduced mass $\mu \equiv m(S)/2$ and solve it numerically (cf.\ \cite{Bicudo:2012qt} for details). While there is strong indication for a bound state in the light scalar isosinglet channel, there seems to be no binding for the light vector isotriplet, or when $q=s$ or $q=c$. To quantify these statements, we list in Table~\ref{binding} the factor by which the reduced mass $\mu$ has to be multiplied to obtain a bound state with confidence level $1 \, \sigma$ and $2 \, \sigma$, respectively (the factors $\leq 1.0$ in the light scalar isosinglet indicate binding). These results clearly show that meson-meson bound states are more likely to exist for $B$ mesons than for $B_s$ or $B_c$ mesons. In other words it seems to be essential for a tetraquark to have both heavy quarks (leading a large reduced mass $\mu$) and light quarks (resulting in a deep and wide potential).

\begin{table}[htb]
\begin{center}
\begin{tabular}{l|cc|cc|cc}
\br
flavor & \multicolumn{2}{c}{light} & \multicolumn{2}{|c}{strange} & \multicolumn{2}{|c}{charm}\\
\mr
confidence level for binding & $1 \, \sigma$ & $2 \, \sigma$ & $1 \, \sigma$ & $2 \, \sigma$ & $1 \, \sigma$ & $2 \, \sigma$ \\
\mr
scalar isosinglet & 0.8 & 1.0 & 1.9 & 2.2 & 3.1 & 3.2 \\
vector isotriplet & 1.9 & 2.1 & 2.5 & 2.7 & 3.4 & 3.5 \\
\br
\end{tabular}
\end{center}
\caption{\label{binding}Factors, by which the reduced mass $\mu = m(S)/2$ in Schr\"odinger's equation has to be multiplied to obtain a four-quark bound state with confidence level $1 \, \sigma$ and $2 \, \sigma$, respectively.}
\end{table} 

% ********************

\subsection{$\bar{Q}Q\bar{q}q$ potentials}

At the moment there are only preliminary results for $\bar{Q}Q\bar{q}q$ potentials corresponding to isospin $I=1$ and $q = c$, i.e.\ $\bar{q}q = (\bar{c}_1 c_2 - \bar{c}_2 c_1) / \sqrt{2}$. Interestingly we observed that all these potentials are attractive, while in the $\bar{Q}\bar{Q}qq$ case only half of them are attractive and the other half is repulsive. This can be understood in a qualitative way by comparing the potential of $\bar{Q}Q$ and of $\bar{Q}\bar{Q}$ generated by one-gluon exchange. For $\bar{Q}\bar{Q}$ the Pauli principle applied to $qq$ implies either a symmetric (sextet) or an antisymmetric (triplet) color orientation of the static quarks corresponding to a repulsive or attractive interaction, respectively. For $\bar{Q}Q$ no such restriction is present, i.e.\ all channels contain contributions of the attractive color singlet, which dominates the repulsive color octet.

$I=0$ requires the computation of an additional diagram and $u/d$ and $s$ quarks are more demanding with respect to HPC resources than $c$ quarks. We expect corresponding results to be available soon.

% ********************
% ********************
% ********************

\section{Conclusions}

We have obtained insights regarding the quark mass dependence of $\bar{Q}\bar{Q}qq$ potentials, which suggest that tetraquark states with two heavy $\bar{b}$ antiquarks seem to be more likely to exist, when there are also two light $u/d$ quarks involved but not $s$ or $c$ quarks.

Preliminary results for $\bar{Q}Q\bar{q}q$ potentials indicate that there are only attractive channels, which is in contrast to the $\bar{Q}\bar{Q}qq$ case.

% ********************
% ********************
% ********************

\section*{Acknowledgments}

We thank Joshua Berlin, Owe Philipsen, Annabelle Uenver-Thiele and Philipp Wolf for helpful discussions. M.W.\ acknowledges support by the Emmy Noether Programme of the DFG (German Research Foundation), grant WA 3000/1-1. This work was supported in part by the Helmholtz International Center for FAIR within the framework of the LOEWE program launched by the State of Hesse.

% ********************
% ********************
% ********************

\end{document}